\documentclass[fleqn,10pt]{wlscirep}
\usepackage[utf8]{inputenc}
\usepackage[T1]{fontenc}

\usepackage{graphicx}
\usepackage{color}
\usepackage{dcolumn}
\usepackage{bm}
\usepackage{url}
\usepackage[english]{babel}
\usepackage{subfigure}
\usepackage{amsfonts}
\usepackage{amsmath}
\usepackage{amssymb}

\def\dd{\mathrm{d}}
\def\ii{\mathrm{i}}

\title{Potential energy of complex networks: a novel perspective}

\author[1,2,+]{Nicola Amoroso}
\author[3,+]{Loredana Bellantuono}
\author[3,2,*]{Saverio Pascazio}
\author[2]{Angela Lombardi}
\author[2]{Alfonso Monaco}
\author[2]{Sabina Tangaro}
\author[3,2]{Roberto Bellotti}

\affil[1]{Dipartimento di Farmacia-Scienze del Farmaco, Universit\`a degli studi di Bari ``A. Moro'', I-70125 Bari, Italy}
\affil[2]{Istituto Nazionale di Fisica Nucleare, Sezione di Bari, I-70126 Bari, Italy}
\affil[3]{Dipartimento Interateneo di Fisica ``M. Merlin'', Universit\`a degli studi di Bari ``A. Moro'', I-70126 Bari, Italy}

\affil[*]{saverio.pascazio@ba.infn.it}

\affil[+]{these authors contributed equally to this work}


\begin{abstract}
We present a novel characterization of complex networks, based on the potential of an associated Schr\"odinger equation. The potential is designed so that the energy spectrum of the Schr\"odinger equation coincides with the graph spectrum of the normalized Laplacian. Crucial information is retained in the reconstructed potential, which provides a compact representation of the properties of the network structure. 

The median potential over several random network realizations is fitted via a Landau-like function, and its length scale is found to diverge as the critical connection probability is approached from above. The ruggedness of the median potential profile is quantified using the Higuchi fractal dimension, which displays a maximum at the critical connection probability. 
This demonstrates that this technique can be successfully employed in the study of random networks, as an alternative indicator of the percolation phase transition. 

We apply the proposed approach to the investigation of real-world networks describing infrastructures (US power grid). Curiously, although no notion of phase transition can be given for such networks, the fractality of the median potential displays signatures of criticality. We also show that standard techniques (such as the scaling features of the largest connected component) do not detect any signature or remnant of criticality.
\end{abstract}
\begin{document}

\flushbottom
\maketitle

\section{Introduction}
\label{sec:intro}
Complex systems, such as political, biological, and financial ones, consist of many elements, whose connections display highly structured patterns \cite{newman2003structure,palla2005uncovering,sporns2011human}. 
Interestingly, some key features, such as preferential attachment, appear to be very general and are observed in very diverse networks \cite{newman2003structure}. Recent investigations have explored the possibility that hidden similarities (and important differences) between networks can be unveiled by analyzing network spectra: the spectral domain can indeed reveal properties which would otherwise remain undetected \cite{farkas2001spectra,nadakuditi2012graph}. 

A number of studies have unearthed interesting relations between the network spectral properties and connectivity \cite{goh2001universal,vukadinovic2001spectral}. In particular, the degeneracy of the lowest eigenvalue of the graph Laplacian associated with the network is equal to the number of its connected components \cite{fiedler1973algebraic}. Results in this field encourage the research of novel spectrum-based frameworks to capture similar patterns in networks of various nature \cite{de2016spectral}, following a recent tendency to explore new tools for network comparison \cite{onnela2012taxonomies,klimm2014resolving,tantardini2019}. Several applications can be envisaged, ranging from the possibility to characterize different information patterns \cite{de2016physics} to the reduction of the structure and complexity of biological, transportation, and social multiplex networks \cite{de2015structural,amoroso2018,amoroso2019}.

In this article we propose a novel approach to characterize complex networks based on the Laplacian spectrum. We associate with the network a one-dimensional Schr\"odinger equation whose eigenvalues coincide with those of the graph spectrum. The potential that appears in such equation is reconstructed through dressing transformations \cite{ramani1995fractal,PhysRevLett.69.398,bittanti1991,wu1990gaussian,van2003riemann}, and provides a compact representation of the network properties, in particular those related to connectivity.
We shall see that the application of quantum-inspired techniques to the study of complex networks turns out to be fecund. On one hand, it offers different perspectives, on the other hand, it is able to capture features that usual methods do not detect.

To test the effectiveness of this new tool, we apply it to a well-known testbed in complex network theory, provided by Erd\"os and R\'enyi (ER) \cite{erdos1959random}. Besides their historical role \cite{bollobas1998random}, random graphs like the ER model are currently used to provide a description of real phenomena such as epidemiological cases \cite{kretzschmar1996measures}, collaboration networks \cite{newman2001random} and social networks \cite{newman2002random}, at least as benchmarks. Moreover, ER networks are characterized by an interesting phase transition, related to the emergence of a giant component, at a critical value of the connection probability between pairs of nodes \cite{barabasi2016network}. We will show that our analysis tools, based on the reconstructed potentials, are able to capture the singular behavior of the network close to the transition. Specifically, three indicators of such criticality will be identified: the length scale, depth and Higuchi Fractal dimension (HFD) \cite {Higuchi1988} of the pointwise median potential, computed on several realizations of the ER network with the same size and connection probability. Finally, to check the validity of the proposed approach in real-world systems, we shall reconstruct potentials from the graph spectra of publicly available complex networks describing infrastructures (US power grid) \cite{konect:2016}. These are real networks, for which no notion of phase transition can be defined. However, interestingly, we shall find that the fractality of the median potential displays signatures of criticality.

The content of this Article is organized as follows: in Section \ref{sec:meth} we present the properties of the Laplacian spectrum and discuss the method for reconstructing the associated potentials via dressing transformations; in Section \ref{sec:ER} we examine the description of the ER network provided by the reconstructed potential framework, focusing on the critical behavior of its length scale, depth and HFD at the phase transition; in Section \ref{sec:rwnet} we use this approach to investigate a real-world network, the US power grid. The details of the dressing transformation methods for the potential reconstruction from the graph spectra are presented in the Supplementary Information.

\section{Setting up the problem: from graph spectra to reconstructed potentials}
\label{sec:meth}
\begin{figure}
\centering
\includegraphics[width=\linewidth]{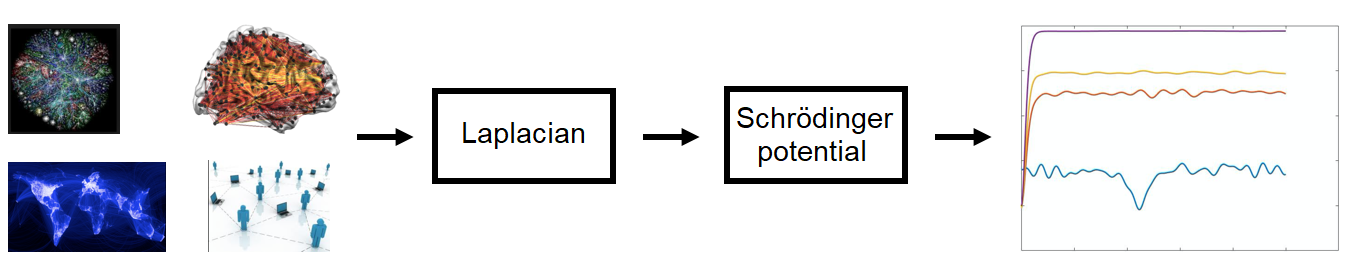}
\caption{Conceptual workflow of the proposed methodology: starting from a given network, one first computes the Laplacian and its spectrum; the potential of a 1D Schr\"odinger equation, yielding the given set of eigenvalues, is then obtained and used to characterize the whole network and investigate its connectivity.}
\label{fig:1}
\end{figure}

Algebraic graph theory is a branch of graph theory devoted to the investigation of graph connectivity properties using results and methods from algebra. In this respect, one of the most interesting results concerns the possibility of infering a connectivity measure by looking at particular eigenvalues of the graph Laplacian \cite{Newman2010}. However, more general properties of a graph could be unearthed by examining all the eigenvalues and the graph energy, defined as their sum \cite{gutman2001energy,gutman2006laplacian}. This approach could result in a substantially novel perspective on the problem of measuring the robustness of real-world complex networks. In the attempt to give a compact representation of the information contained in the whole graph spectrum, we apply a methodology, based on non-linear equations, to retrieve a one-dimensional potential given a set of energy levels (see Fig. \ref{fig:1} for a schematic overview). 

Accordingly, starting from a generic (unweighted and undirected) network, we first compute its Laplacian and  eigenvalues, and then reconstruct the unique Schr\"odinger potential associated with the spectrum. The potential profile will yield a snapshot of the network configuration, providing a novel perspective on the study of its connectivity.

\subsection{Laplacian spectrum}
\label{sec:Lspectrum}

We introduce here the formalism that is necessary to analyze complex networks in the framework of graph theory \cite{Newman2010} and recall a few basic notions that will be useful in our analysis. A graph $\mathcal{G}=(\mathcal{N},\mathcal{E})$ is defined through a set $\mathcal{N}$ of $N$ nodes and a set $\mathcal{E}$ of edges connecting them. The adjacency matrix $A$ of $\mathcal{G}$ is a matrix whose elements $a_{ij}$ are nonvanishing only if a connection between node $i$ and node $j$ exists. 
In general, graphs can be built by assigning a weight and an orientation to each edge. In the present work, we shall focus on undirected and unweighted graphs, whose adjacency matrices are symmetric and binary (i.e., consisting only of $0$ and $1$ elements). Moreover, for the sake of simplicity, the networks considered in this work will not include loops, namely links connecting a node to itself. For each graph $\mathcal{G}$, given its adjacency matrix $A=\{a_{ij}\}$, the number of connections of each node, namely the node degree $d_i$, is simply calculated by summing column- or row-wise the adjacency matrix. Accordingly, one defines the degree matrix $D$ as the diagonal matrix with $D_{ii}=d_i$. The Laplacian $L = \{L_{ij}\}$ of $\mathcal{G}$ is defined as the difference  $D-A$, so that:
\begin{equation}
L_{ij} = \left\{ \begin{array}{rl}
 d_i &\mbox{ if $i=j$} \\
  -1 &\mbox{ if $i,j$ adjacent} \\
  0 &\mbox{ otherwise} \\
       \end{array} \right. .
\end{equation}
The normalized Laplacian $\mathcal{L} = \{\mathcal{L}_{ij}\}$ of $\mathcal{G}$ is defined as the matrix with elements
\begin{equation}\label{eq:normalizedLaplacian}
\mathcal{L}_{ij} = \left\{ \begin{array}{rl}
 1 &\mbox{ if $i=j$ and $d_i \neq 0$ } \\
  -\frac{1}{\sqrt{d_i d_j}} &\mbox{ if $i,j$ adjacent} \\
  0 &\mbox{ otherwise} \\
       \end{array} \right.
\end{equation}
which can be expressed as $\mathcal{L}=D^{-1/2}LD^{1/2}$, with the convention $\left(D^{-1/2}\right)_{i,i}=0$ for $d_i=0$, namely if the node associated with index $i$ is isolated. If the complex network features components that are disconnected from each other, both the $L$ and $\mathcal{L}$ matrices can be recast into block-diagonal forms, with each block corresponding to a specific component. The spectrum of the normalized Laplacian, which is also called the spectrum of the graph, provides comprehensive information on the structure of the network, with regard to the number of its connected components and their dimensions. The eigenvalues $\lambda_1 \leq \lambda_2 \leq \dots \leq \lambda_N$ of $\mathcal{L}$ satisfy $0\leq \lambda_{i} \leq 2$. In particular, $\lambda_{1}=0$ is always an eigenvalue, whose multiplicity coincides with the number of connected components in the network. The eigenvalue $\lambda_2$, called algebraic connectivity or Fiedler eigenvalue, is therefore nonvanishing if and only if the network is connected \cite{fiedler1973algebraic}. To give an idea of the spectra of $\mathcal{L}$ corresponding to peculiar structures, it is worth reviewing a few notable cases which will be relevant for our analysis \cite{Chung1997}:
\begin{itemize}
    \item disconnected graph ($A_{ij}=0$ for all $i,j$): $0$ is the only eigenvalue, with multiplicity $N$;
    \item complete graph $K_N$ ($A_{ij}=1$ for all $i\neq j$): the eigenvalues are $0$, with multiplicity $1$, and $N/(N-1)$, with multiplicity $N-1$;
    \item path $P_N$ on $N$ vertices: the eigenvalues are $1-\cos\frac{\pi k}{N-1}$, with $k=0,\dots,N-1$;
    \item cycle $C_N$ on $N$ vertices: the eigenvalues are $1-\cos\frac{2\pi k}{N}$, with $k=0,\dots,N-1$.
\end{itemize}
If the network includes components $K_n$, $P_n$, $C_n$ with $n<N$, the eigenvalues of the $n\times n$ Laplacians associated with such components will contribute to the spectrum of the whole network Laplacian \eqref{eq:normalizedLaplacian}.

\subsection{Reconstructing potentials through dressing transformations}
We now associate to a given network a potential and a 1D Schr\"{o}dinger equation, whose energy levels coincide with the spectrum of the normalized Laplacian of the network. For this purpose, we shall apply a method based on the dressing transformation, proposed in \cite{PhysRevLett.69.398} and employed e.g. in \cite{ramani1995fractal,van2003riemann}.

For a given network, we compute the eigenvalues $\lambda_1 \leq \lambda_2\leq \dots\leq \lambda_N$ of the normalized Laplacian $\mathcal{L}$, and consider the shifted spectrum 
\begin{equation}
E_n = \lambda_n - \lambda_N ,
\label{eq:En}
\end{equation}
whose values are in $[-2,0]$. Hence, we derive the potential $V(x)$ such that all the $\widetilde{N} < N$ nonvanishing shifted eigenvalues are energy levels of the Schr\"odinger equation ($\hbar=1, m=1/2$)
\begin{equation}\label{eq:schro}
-\partial_x^2 \psi(x) + V(x) \psi(x) = E_n \psi(x),
\end{equation}
with $\psi(x)$ a normalizable wavefunction. We obtain the potential $V$ by applying a method, whose details are outlined in the Supplementary  Information, that is based on the iteration of a two-step procedure: we first solve the Riccati equation
\begin{equation}
\label{eq:diff}
\displaystyle f'_n(x) - f_n^2(x) + V_{n+1}(x) - E_n = 0 \quad \text{with } f_n(0)=0\,,
\end{equation}
and then update the potential according to
\begin{equation}
V_n(x) = V_{n+1}(x) + 2 f_n'(x).
\end{equation}
The iteration starts at $n=\widetilde{N}$, with $V_{\widetilde{N}+1}=0$ and proceeds until $n=1$, yielding the final potential $V_1$ that can be identified with $V$ in Eq. \eqref{eq:schro}. Notice that the potentials obtained by the above procedure are even by construction.

The sequence of equations (\ref{eq:diff}) can be integrated with any standard numerical method. Following the approach proposed in \cite{van2003riemann}, we adopted a fourth-order Runge-Kutta (RK) method with step size $h = 10^{-5}$ to guarantee a high reconstruction accuracy.
It is worth remarking that the dressing transformation is not the only technique to solve the inverse problem of finding the potential corresponding to a given set of levels. Another possible choice is the variational method described in Ref. \cite{wu1990gaussian}, which however requires a search in a very large parameter space. Additional details are given in the Supplementary Information.

\section{Erd\"{o}s-R\'{e}nyi phase transition in the reconstructed potential framework}
\label{sec:ER}
A \textit{random} network is a set of $N$ nodes, with edges randomly placed among them. The original model of a random network was introduced by Erd\"{o}s and R\'{e}nyi (ER) in their seminal work \cite{erdos1959random}, where they defined an ensemble $G(N,L)$ of graphs having a fixed number $L$ of randomly placed edges. Here we will focus on a different formulation, proposed by Gilbert \cite{Gilbert1959} and characterized by the fact that the number of edges is not fixed. This model is based on an ensemble of networks $G(N,p)$, in which each node pair is connected with probability $p$. A parameter to quantify the connectivity for networks in this ensemble is the \emph{average degree}
\begin{equation}
\label{eq:kvsp}
\langle k \rangle = p(N-1). 
\end{equation}
Erd\"{o}s and R\'{e}nyi predicted the existence of a phase transition (in the thermodynamical limit) at $\langle k \rangle = 1$, corresponding to the critical probability $p_c=1/(N-1)$. This transition consists in the appearance of percolation and is related to a change in the scaling behavior of the typical dimension $N_{LCC}$ of the largest connected component in each graph of the ensemble, which becomes a \textit{giant component} at the critical probability. In the subcritical regime ($p<p_c$), $N_{LCC}$ is of order $\ln N$, while in the supercritical one ($p>p_c$), $N_{LCC}$ scales like the number of nodes $N$. At criticality, $N_{LCC}\sim N^{2/3}$. 
\cite{erdos1959random, Gilbert1959,barabasi2016network}

\subsection{Reconstructed potentials and connection probabilities}

We consider different realizations of the ER complex network, with $N=500$ nodes and different values of the connection probability $p$. To get a qualitative understanding of the relation between the network structure, the graph spectrum and the reconstructed potential, let us first look at three particular cases: $p=10^{-4}$, $p=0.9$ and the critical probability $p=p_c=1/(N-1)\simeq 2 \times 10^{-3}$. 

\begin{figure}
\centering
\includegraphics[scale=0.29]{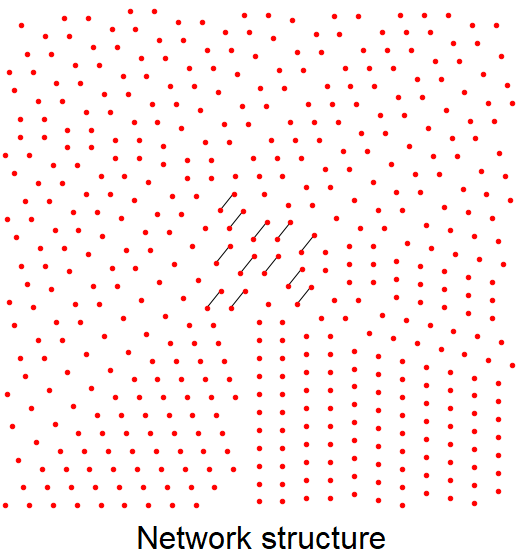}
\includegraphics[scale=0.23]{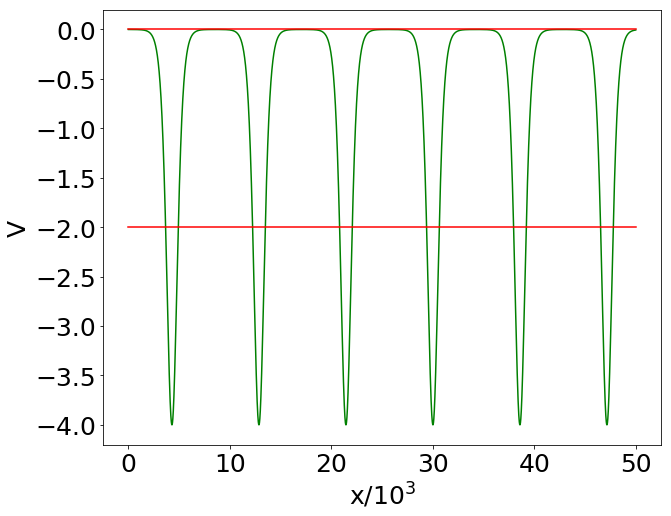}
\includegraphics[scale=0.23]{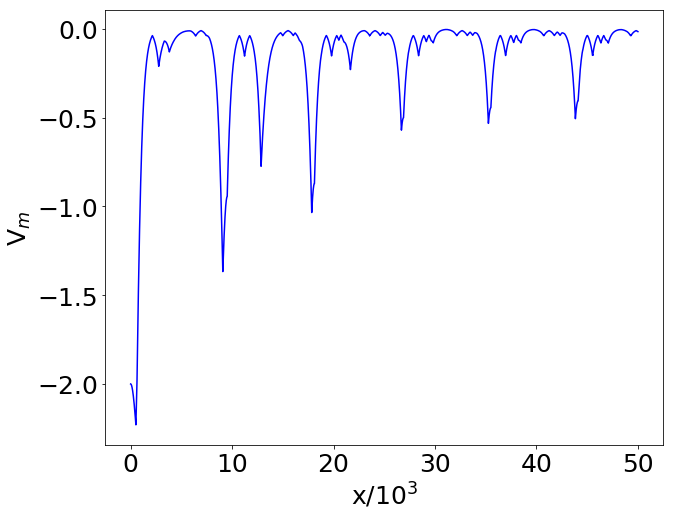}
\caption{Left panel: complex network with $N=500$ nodes, generated using the ER model with connection probability $p=10^{-4}$. Central panel: shifted graph spectrum $E_n$ (red horizontal lines) and reconstructed potential $V(x)$ (green) corresponding to the network in the left panel, with $x$ the dimensionless variable of the Schr\"{o}dinger equation \eqref{eq:schro}; since $V(x)$ is an even function, only the positive $x$ semi-axis is displayed. Right panel: pointwise median potential $V_m(x)$ obtained from the reconstructed potentials of $M=100$ network realizations with the same $N$ and $p$.}
\label{fig:ER_low_p}
\end{figure}

The statistical ensemble for $p=10^{-4}$ is made of almost disconnected networks. In each realization, a few pairs of connected nodes are present, and connected components with more than two nodes are extremely rare. Thus, the typical spectrum $\lambda_n$ ($n=1,\dots,N$) of $\mathcal{L}$ consists of the highly degenerate eigenvalue $0$, whose multiplicity coincides with the number of connected components, and the eigenvalue $2$, related to the presence of paths $P_{2}$ of two nodes (see the examples in Section \ref{sec:Lspectrum}), which are on average $pN(N-1)/2\simeq 12.5$. These eigenvalues correspond to the values $-2$ and $0$ of the shifted spectrum \eqref{eq:En}, respectively. This structure of the spectrum tends to appear in the vast majority of realizations, one of which is displayed in Fig.\ \ref{fig:ER_low_p}. 
We observe that the number of wells in the reconstructed potential tends to coincide with the ground state degeneracy, which, in this case, is equal to $488$, while the excited state has degeneracy $12$. The similarity of the reconstructed potentials is reflected in the regularity of the pointwise median potential $V_m(x)$ (average over $M=100$ network realizations), shown in the right panel of Fig.\ \ref{fig:ER_low_p}.

\begin{figure}
\centering
\includegraphics[scale=0.30]{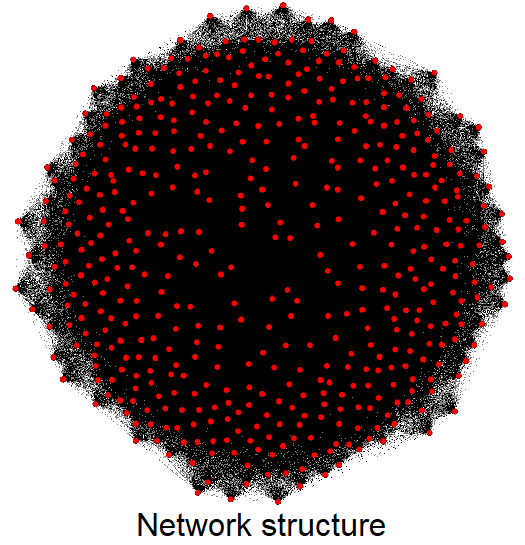}
\includegraphics[scale=0.23]{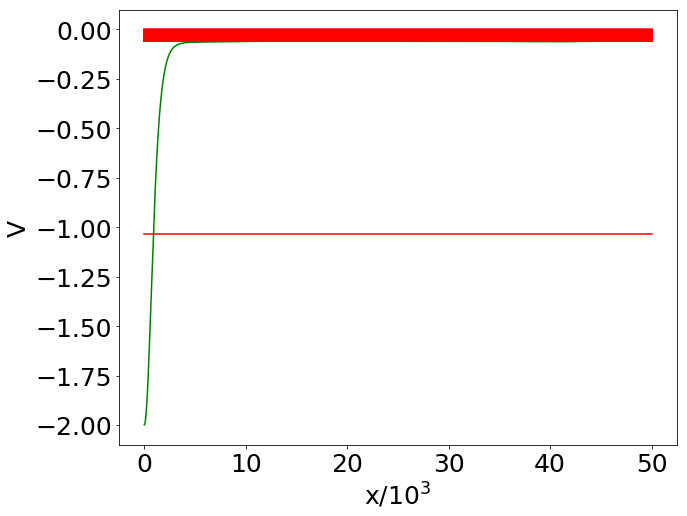}
\includegraphics[scale=0.23]{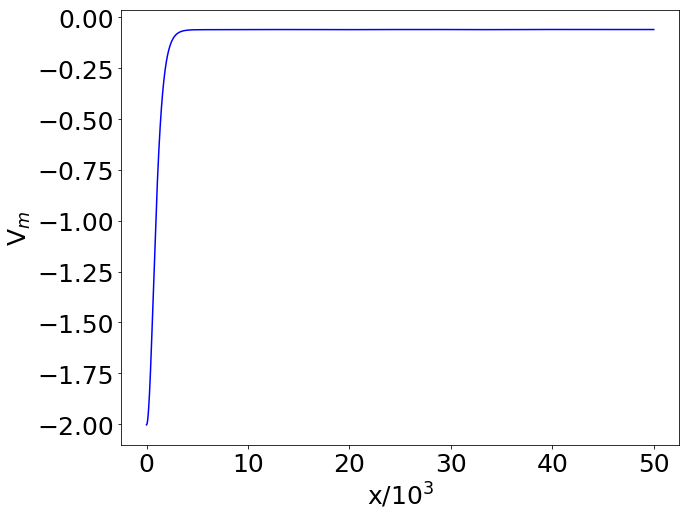}
\caption{Left panel: complex network with $N=500$ nodes, generated using the ER model with connection probability $p=0.9$. Central panel: shifted graph spectrum $E_n$ (red horizontal lines) and reconstructed potential $V(x)$ (green, only positive $x$ displayed) corresponding to the network in the left panel, with $x$ the dimensionless variable of the Schr\"{o}dinger equation \eqref{eq:schro}. Right panel: pointwise median potential $V_m(x)$ over $M=100$ network realizations with the same $N$ and $p$.}
\label{fig:ER_high_p}
\end{figure}

For $p=0.9$, in basically all network realizations the graph consists of a single component, with a high density of links. In such configuration, the spectrum $\lambda_n$ of $\mathcal{L}$ consists of a nondegenerate eigenvalue $0$, separated by a gap from the other eigenvalues, that concentrate around $1$. Indeed, when the connection probability is close to $p=1$, the spectrum approximates that of the complete graph $K_{N}$. Also in this case the structure of the spectrum is weakly dependent on the specific network realization, a feature which is again reflected in the similarity of the reconstructed potentials. The typical potential profile is characterized by a single minimum around $x=0$, and rapidly increases approaching an almost constant value. The statistical variability of the reconstructed potentials is extremely low, leading to a very smooth median $V_m(x)$ (average over $M=100$ network realizations), manifest in Fig.\ \ref{fig:ER_high_p}.

\begin{figure}
\centering
\includegraphics[scale=0.29]{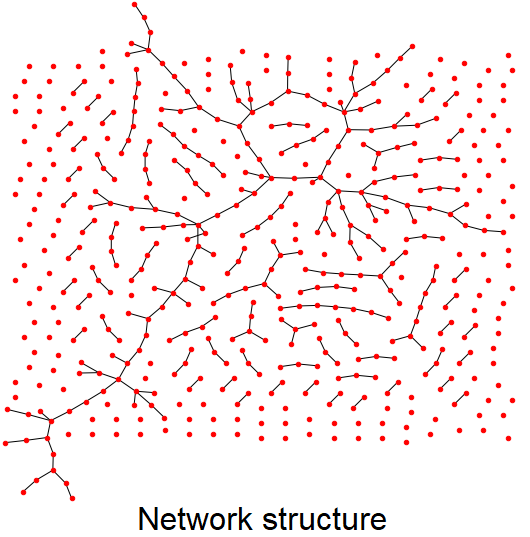}
\includegraphics[scale=0.23]{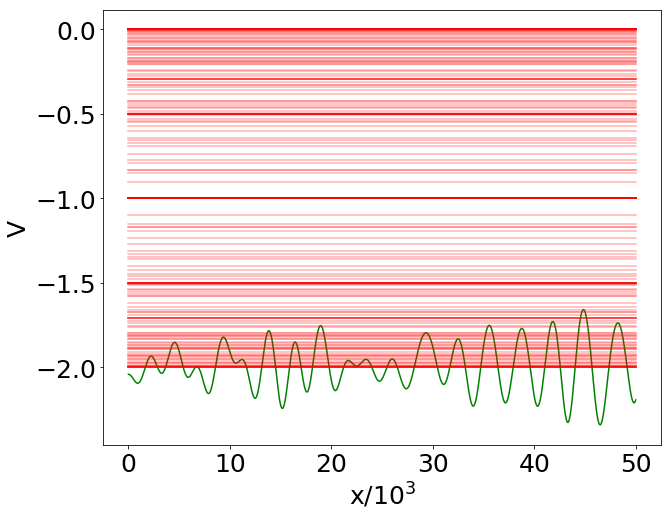}
\includegraphics[scale=0.23]{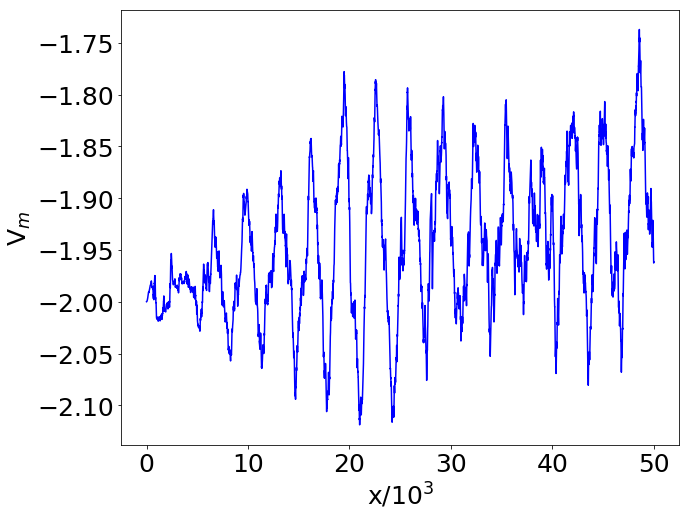}
\caption{Left panel: complex network with $N=500$ nodes, generated using the ER model with connection probability $p=p_c=1/(N-1)\simeq 2\cdot 10^{-3}$. Central panel: shifted graph spectrum $E_n$ (red horizontal lines) and reconstructed potential $V(x)$ (green, only positive $x$ displayed) corresponding to the network in the left panel, with $x$ the dimensionless variable of the Schr\"{o}dinger equation \eqref{eq:schro}. Right panel: pointwise median potential $V_m(x)$ over $M=100$ network realizations with the same $N$ and $p$.}
\label{fig:ER_critical_p}
\end{figure}

At the critical probability $p=p_c=1/(N-1)$, (almost) each realization of the network is characterized by the emergence of a giant component, involving a number of nodes of order $N^{2/3}$.\cite{erdos1959random, Gilbert1959,barabasi2016network} The remaining nodes are either connected into smaller size components or isolated. The eigenvalues $\lambda_n$ of the graph spectrum are distributed in the full interval $[0,2]$, with a larger concentration around the endpoints.
This behavior is mainly due to the presence of the giant component, which can be approximately described as a path, that contains also small ramifications and cycles. For this reason the contribution of the giant component to the graph spectrum is qualitatively similar to the spectrum of a path $P_{n}$ with $n\sim N^{2/3}$. Although the giant component displays general (and common) features in all the network realizations at critical $p$, the details of its nontrivial structure can hardly be reproduced. Therefore, since giant components corresponding to different realizations are generally characterized by very different patterns and micro-structures, and even different dimensions, the shifted eigenvalues $E_n$ (positions of the associated spectral lines in the central panel of Fig.\ \ref{fig:ER_critical_p}) will be wildly fluctuating. Smaller components and isolated nodes in the network will contribute to the spectrum of $\mathcal{L}$ with sparse and degenerate eigenvalues, as in the case of low $p$. In Fig.\ \ref{fig:ER_critical_p} we show the reconstructed potential for a single realization, whose eigenvalues $E_n$ correspond to the shifted spectrum \eqref{eq:En} lying between $-2$ and $0$. The shape of this potential, characterized by irregular oscillations around a constant value, with no appreciable increase in the considered $x$ range, is actually rather similar for all realizations at critical $p$. However, the features of these oscillations wildly differ for different realizations, since they are subject to the same variability that characterizes the spectrum associated with the giant component. As a result, the profile of the median $V_m(x)$ is very irregular. This was observed to be true also for a surprisingly small number of realizations. This observation is central: we shall argue that the irregularity of the average potential at criticality is but a manifestation of the emergence of fractality. 

\begin{figure}
\centering
\includegraphics[width=0.4\linewidth]{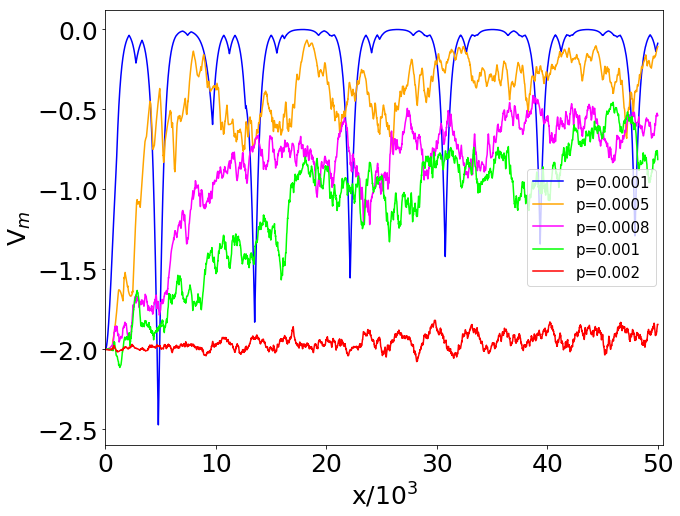}
\includegraphics[width=0.4\textwidth]{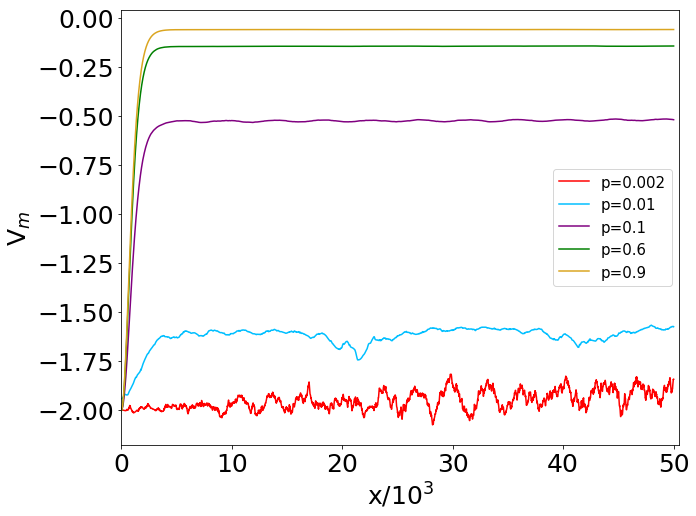}
\caption{Median $V_m(x)$ of the reconstructed potentials $V_i(x)\,(i=1,\dots,M)$, for $M=100$ ER networks, $N=500$ nodes and connection probability $p$. The critical value is $p_c = 1/(N-1)\simeq 2 \times 10^{-3}$.
Left panel: $p<p_c$, with $p$ ranging from $10^{-4}$ (top) to $2 \times 10^{-3}$ (bottom). 
Right panel: $p>p_c$, with $p$ ranging from $2 \times 10^{-3}$ (bottom) to $0.9$ (top). 
The values of $p$ are given in the insets. 
For $p < p_c$, the median potentials are not smooth and decrease as $p$ approaches the critical value from below (left). For $p\gg p_c$, $V_m(x)$ rapidly increases and reaches a saturation value (right). 
}
\label{fig:3}
\end{figure}

\subsection{Length scale of the reconstructed potential}

The median potentials $V_m(x)$ obtained from $M=100$ realizations of the ER network with $N=500$ nodes, for different values of $p$, are shown in Fig.\ \ref{fig:3}. One observes a very different behavior, depending on whether $p$ is smaller or larger than $p_c$. For $p<p_c$, $V_m$ features irregular oscillations superposed to an increasing trend, that becomes less and less steep as criticality is approached. For $p>p_c$, the curves become smoother (differentiable), and a minimum in the origin gradually appears, followed by a rise to an almost constant plateau. As we shall see, this behavior is well approximated by a Landau potential $V(x)=a\,\mathrm{sech}^2(x/b)$, with $-a$ and $b$ denoting the depth and width of the trapping potential, respectively \cite{Landau1981Quantum}.

\begin{figure}
\centering
\includegraphics[scale=0.23]{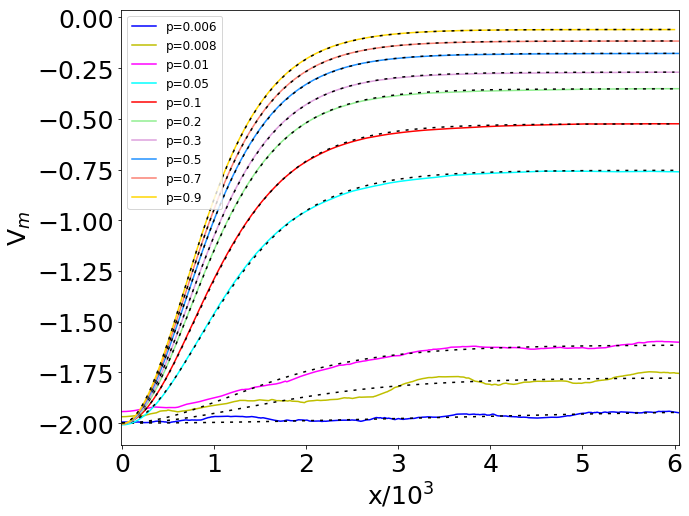} 
\includegraphics[scale=0.23]{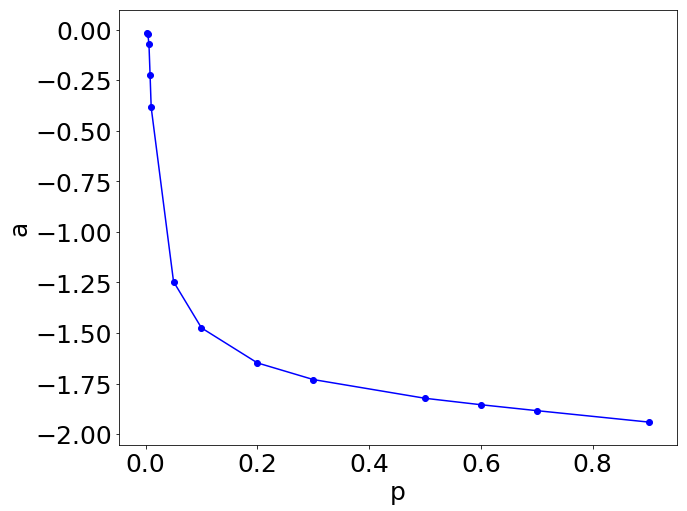}
\includegraphics[scale=0.23]{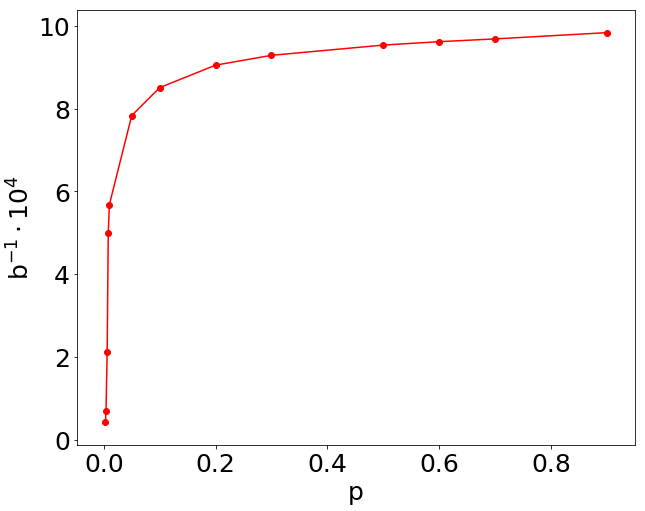}
\caption{Potentials for $p>p_c$. Left: comparison between median potentials $V_m$ in the right panel of Fig.\ \ref{fig:3} (continuous lines) and their best fits with the functional form $V_\textrm{fit}(x,p)$ in Eq.\ \eqref{eq:Vfit} (dashed lines). 
Center: best fit values of $a$ (whose absolute value represents the depth of the potential well) vs $p$; the fit yields $a \sim \left(p-p_{c}\right)^{-\alpha}$, with $\alpha \simeq 1.60 \pm 0.22$. Right: best fit values of $b$ (width of the potential well) vs $p$; the fit yields $b \sim\left(p-p_{c}\right)^{-\beta}$, with $\beta\simeq 1.35 \pm 0.13$. 
}
\label{fig:fit_V_median}
\end{figure}

The reason for the dependence of $V_m(x)$ on $p (>p_c)$ becomes evident when one compares the shifted Laplacian spectra in the central panels of Figs.\ \ref{fig:ER_high_p} and \ref{fig:ER_critical_p}: by approaching the percolation transition from above, the reconstructed potentials tend to become steeper, in order to accommodate a larger number of nondegenerate eigenvalues in the upper part of the spectrum. This observation suggests a fit of the median potential $V_{m}(x)$ with an expression that interpolates between the large-$p$ and critical-$p$ regimes: we consider an ansatz of the following type 
\begin{equation}
\label{eq:Vfit}
    V_\textrm{fit}(x,p)= a(p)\,\mathrm{sech}^2\left(\frac{x}{b(p)} \right)-2-a(p)\,,
\end{equation}
in the attempt to fit the slower increase of $V_{m}(x)$ towards its asymptotic (large-$x$) value in the vicinity of the phase transition. 
Figure \ref{fig:fit_V_median} displays the comparison between $V_m(x)$ and $V_\textrm{fit}(x,p)$ at different values of $p>p_c=2 \times 10^{-3}$, and the dependence of the fit parameters $a$ and $b$ on the connection probability $p$ (at $N=500$). The relative errors on the fit parameters $a$ and $b$ are always smaller than $0.02\%$ for $p\leq 5 \times 10^{-2}$, and increase as the  critical point is approached, reaching $1.16\%$ for $a$ and $1.65\%$ for $b$ at $p=3\times 10^{-3}$, where the oscillations of the median potential profile reduce the fit accuracy.  

The median potential captures the critical behavior of the system, as it emerges from the divergence of the fit parameters as $p$ approaches $p_c$ from above. In particular, the scales $a$ and $b$, which set respectively the depth and width of the trapping potential, diverge like
\begin{equation}
a \sim \left(p-p_{c}\right)^{-\alpha}\,, \qquad b \sim \left(p-p_{c}\right)^{-\beta}
\label{eq:abdiv}
\end{equation}
at the percolation transition, with critical exponents $\alpha \simeq 1.60 \pm 0.22$ and $\beta\simeq 1.35 \pm 0.13$. 
This shows that the very structure of the potential of the 1D Schr\"odinger equation, as well as the parameters that characterize it, detect the percolation phase transition of the associated network. 
We shall explore this association in more detail in the next subsection and show that the median potential becomes \emph{fractal} at the phase transition.
A few additional properties of the potential $V_\textrm{fit}$ are discussed in the Supplementary Information.

\subsection{Transition characterization through Higuchi Fractal Dimension}
\label{sec:higfract}

\begin{figure}
\centering
\includegraphics[width=0.5\textwidth]{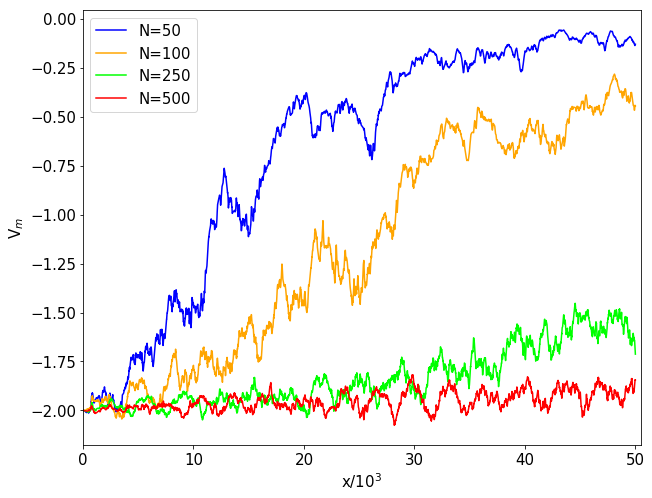}
\caption{Median $V_m(x)$ of the reconstructed potentials $V_i(x)$ $(i=1,\dots,M)$ for $M=100$ ER networks, for different network size $N$, always at the critical probability $p_c = 1/(N-1)$. Each plot is obtained at a fixed value of $N$, specified in the legend.}
\label{fig:crit_pot_vs_N}
\end{figure}

We observed (Fig.\ \ref{fig:3}) that the median potential profile is less smooth at the critical probability $p_c$ than in the low- and high-$p$ regimes, due to the spectral fluctuations of the associated networks. Figure \ref{fig:crit_pot_vs_N} shows, for different values of $N$, the median potentials over $M=100$ realizations of the network, always at the critical connection probabilities $p_c=1/(N-1)$; it can be noticed that the increasing trend in $x$ of the median potential is more manifest for networks of smaller size $N$. At $N=500$, one observes oscillations but no overall increasing trend with $x$.

We now introduce a measure to quantify the lack of smoothness of $V_m$. In Ref.~\cite{ramani1995fractal}, the ruggedness of the potentials reconstructed from given sets of energy levels was characterized in terms of a qualitative measure of fractal dimension. Here, we shall use the Higuchi Fractal Dimension (HFD) \cite{Higuchi1988}, that, having been originally introduced to characterize time series, is especially suited to describe the profile of functions of one variable. From a given sequence $\{F_1,F_2,\dots,F_\nu\}$ one extracts the subsets 
\begin{equation}\label{eq:subseq}
\{F_i, F_{i+k},\dots, F_{i+\left\lfloor \frac{\nu-i}{k} \right\rfloor k} \}, \quad \text{with } i=1,2,\dots, k
\end{equation}
of values corresponding to indices separated by a positive integer $k<\nu$ and starting from $i$. The quantities
\begin{equation}\label{eq:sequenceHFD}
L_{i}(k)=\frac{\nu-1}{\left\lfloor\frac{\nu-i}{k}\right\rfloor}\sum_{j=1}^{\left\lfloor\frac{\nu-i}{k}\right\rfloor}\left\lvert F_{i+jk}-F_{i+(j-1)k} \right\rvert
\end{equation}
represent properly normalized measures of the mean distance between neighboring values in \eqref{eq:subseq}. The terms $L_{i}(k)$ corresponding to the same spacing $k$ can then be averaged over all possible initial points to obtain
\begin{equation}\label{eq:lengthHFD}
\langle L(k) \rangle = \frac{1}{k}\sum_{i=1}^{k}L_i(k)\,.
\end{equation}
If $\langle L(k) \rangle \sim k^{-D}$, the exponent $D$ is called the HFD of the sequence $\{F_1,F_2,\dots,F_\nu\}$. In practice, the above dependence holds only in a certain range of $k$; in our analysis, we have considered $2\leq k \leq 800$.

\begin{figure}
\centering
\includegraphics[scale=0.45]{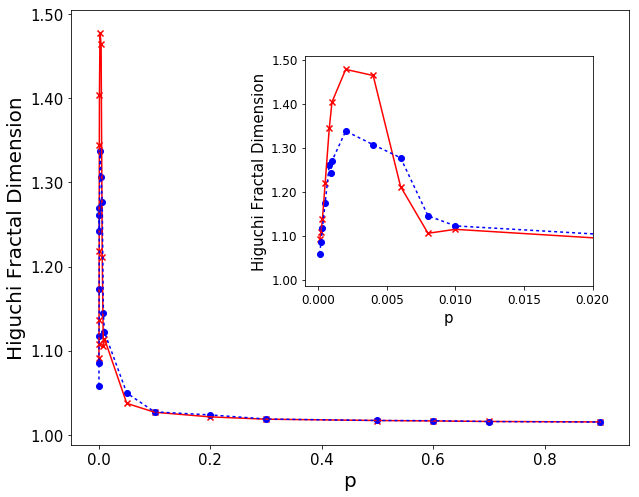}
\caption{Higuchi fractal dimension of the median potentials on a set of $M$ realizations of ER graphs with $N=500$ nodes, as a function of the connection probability $p$. The (blue) dots and dashed curve refer to the median potentials in Fig.\ \ref{fig:3}, obtained from $M=100$; the (red) crosses and full curve show the HFD for $M=1000$.
}
\label{fig:HFD_vs_p}
\end{figure}

Figure \ref{fig:HFD_vs_p} shows the HFD values of the median potentials computed on $M$ realizations of ER networks with $N=500$ nodes at fixed $p$, with $M=100$ (see Fig. \ref{fig:3}) and $M=1000$. From both plots, we observe that the fractal dimension peaks around the critical probability, while it drops to a value close to one in the low- and high-$p$ regimes. The inset in Fig.\ \ref{fig:HFD_vs_p} displays a (close) zoom of the peaks around $p=p_c=1/(N-1)$, showing that the maximum value is attained at the critical probability, for the explored values of $p$ and $N$. 

The maximum of the HFD can be therefore be considered as an indicator of criticality. We observe that the plotted values at each connection probability $p$ depend on the specific ensemble of $M$ randomly sampled realizations of the network. We actually checked that plots from independent sets, each containing $M=100$ realizations, generally fluctuate around a mean curve. However, despite this variability, a general trend emerges, characterized by the presence of a peak around the critical probability, in all the considered sets of $M=100$ realizations. 

\begin{figure}
\includegraphics[scale=0.245]{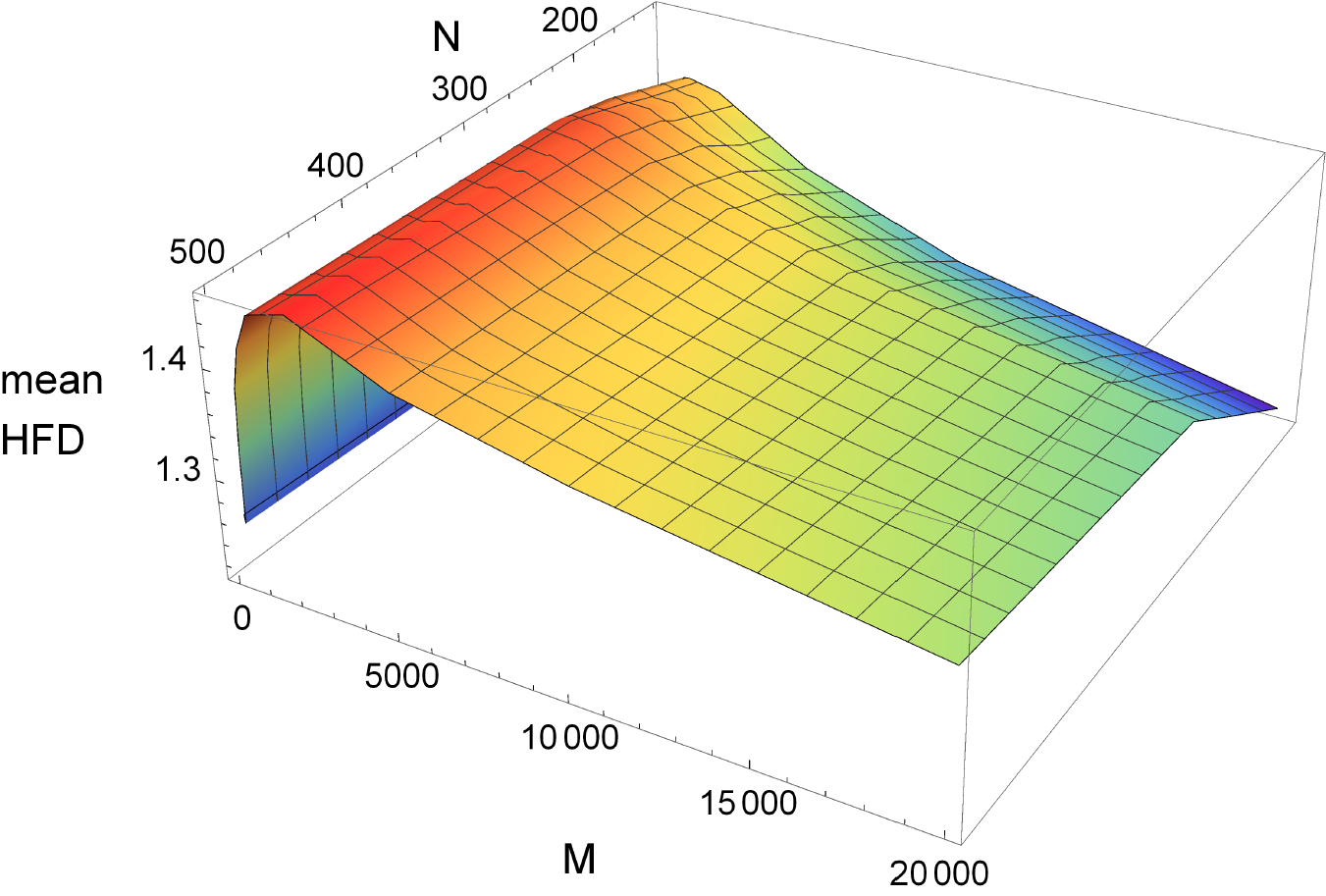}
\hspace{0.2cm}
\includegraphics[scale=0.315]{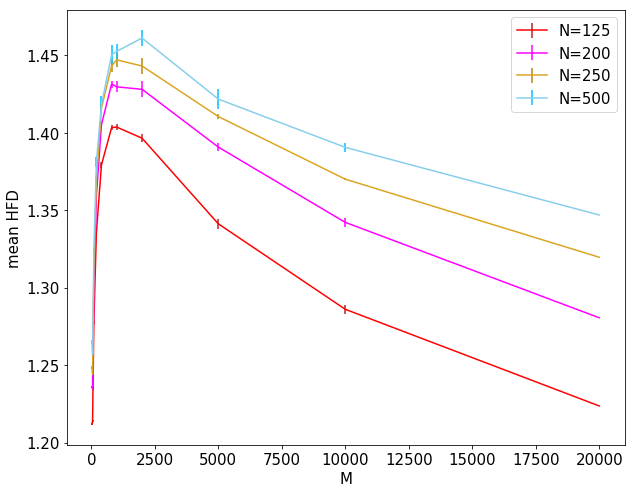}
\caption{Left: mean HFD values at criticality vs $M$ and $N$; Right: error bars for the some data plots. Each distribution contains $s(M)=20000/M$ HFD values of the median potentials on $M$ realizations of ER graphs with $N$ nodes, at the critical connection probability $p=p_c=1/(N-1)$. The position of the maximum is largely independent of $M$.
} 
\label{fig:HFD_vs_M}
\end{figure}

Finally, a comparison between the two plots displayed in Fig.\ \ref{fig:HFD_vs_p} reveals that the values of HFD obtained for different numbers $M$ of realizations are sensibly different when the connection probability approaches the critical value. This result motivates us to systematically investigate the behavior of the fractal dimensions at $p=p_c=1/(N-1)$ for networks with varying numbers of nodes $N$ and realizations $M$ concurring to the median potential. At criticality, for each value of $N\in\{125,200,250,500\}$ we generated $20000$ networks, which were grouped into statistically independent sets of $M\in\{25,50,100,200,400,800,1000,2000,5000,10000,20000\}$ realizations, each with cardinality $s(M)=20000/M$. We then computed the HFD values of the median potentials at criticality, corresponding to sets with the same $M$, obtaining distributions of fractal dimension consisting of $s(M)$ elements. We report in Fig.\ \ref{fig:HFD_vs_M} the mean HFD of such distributions as a function of $M$ and $N$, with the error bars displayed in the bottom panel obtained by dividing the standard deviation of the distribution related to a specific $M$ by $\sqrt{s(M)}$. In all the plots the HFD displays an initial increase, due to the fact that fractality emerges when the median potential is computed over a relevant number of realizations, while the single potentials are not fractal (HFD$\simeq 1$). 
The position of the maximum is largely independent of $M$, and the HFD decreases for large $M$. 
We observe that such a decrease becomes less significant for larger values of $N$. It would be tempting to assume that, as $N$ increases further, the HFD reaches a plateau after the maximum, as a function of $M$. However, our data do not enable us to safely draw this conclusion (even by a tentative analysis of finite-size scaling), nor to determine the precise value of the maximum for $N, M \to \infty$. Additional details on the numerical procedure are outlined in the Supplementary Information.

 \section{Analysis of a real-world network}\label{sec:rwnet}

\begin{figure}
\centering
\includegraphics[scale=0.30]{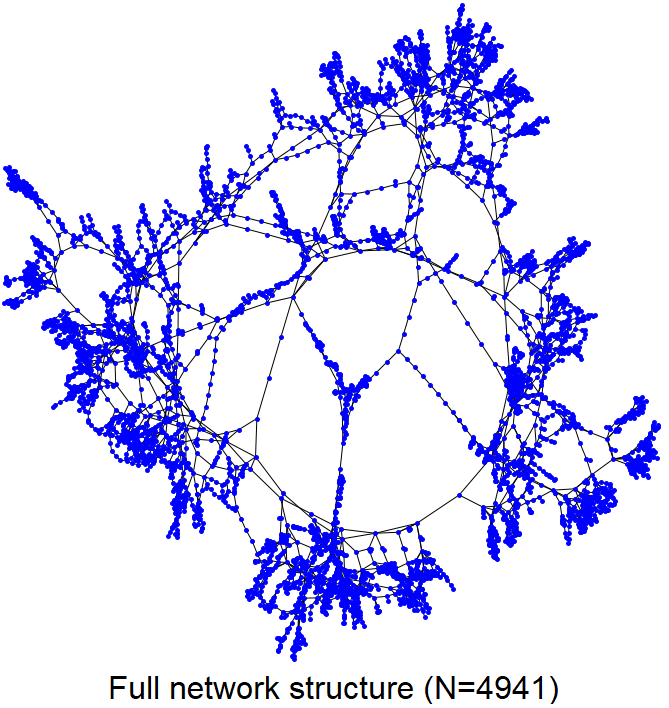}
\includegraphics[scale=0.30]{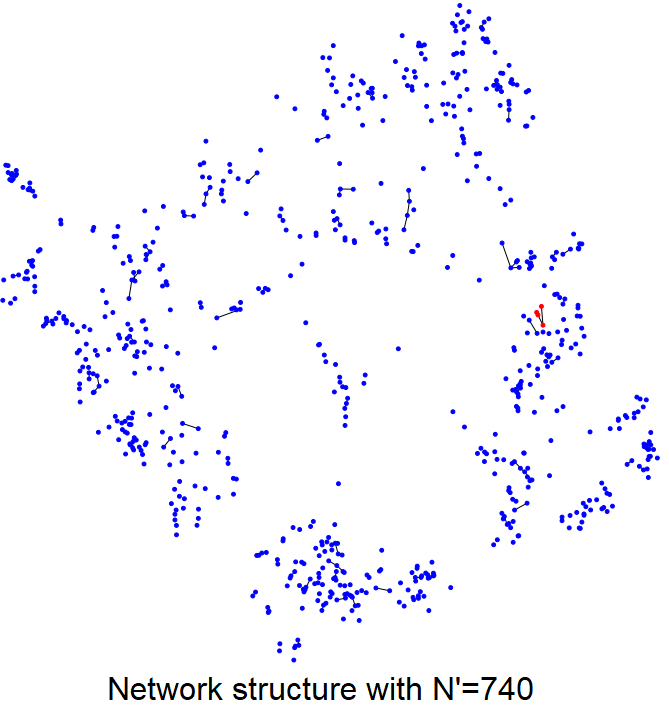}\\
\vspace{0.5cm}
\includegraphics[scale=0.30]{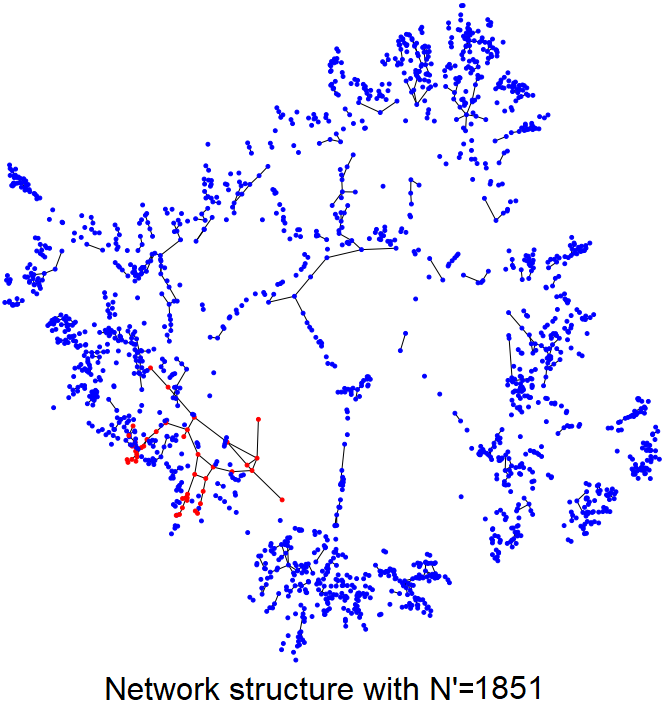}
\includegraphics[scale=0.30]{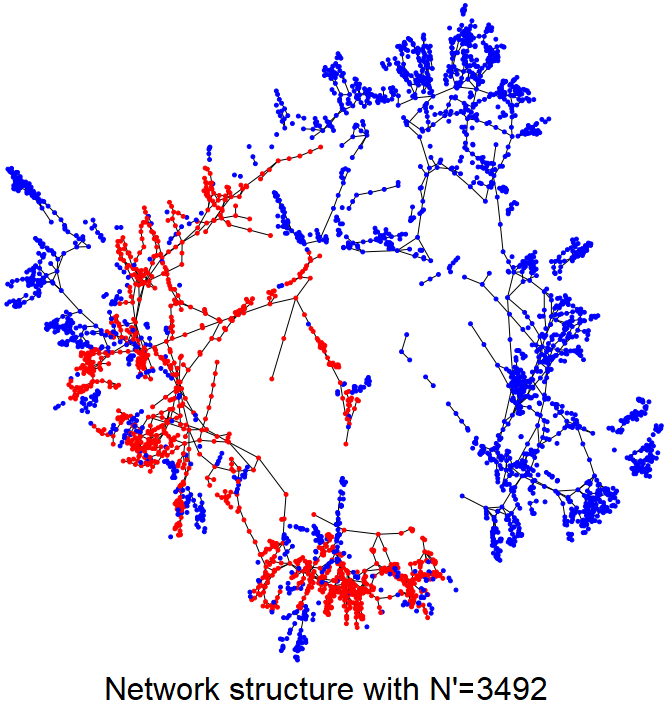}
\caption{Graph representation of the US power grid (top left panel) and realizations of sub-sampled networks with $N'=740$ and $\langle k \rangle = 0.4$ (top right), $N'=1851$ and $\langle k \rangle = 1.01$ (bottom left), $N'=3492$ and $\langle k \rangle = 1.89$ (bottom right); the largest connected component in each sub-sampled network is highlighted in red.
(The layout of the graph is different from that used in Fig.\ \ref{fig:ER_critical_p}.)
} \label{fig:US_power_grid_network}
\end{figure}

The formalism based on the reconstructed potentials and the HFD, developed in the previous sections and used to characterize ensembles of ideal ER networks, will now be applied to the description of a real-world situation. It is important to stress that a real-world network is not a random graph in the sense of ER, and does not enjoy the same idealized features. It is not obvious, for example, that a real-world network percolates, certainly not in the same way as a random graph. For instance, the very presence of a unique giant component and its scaling features cannot be taken for granted. We shall now look at a specific example and consider a number of typical quantities that characterize it.

The data analyzed in this section are taken from a public database \cite{konect:2016} and represent the power grid of the Western States of the USA \cite{watts1998collective}). This system can be modeled as an undirected unweighted complex network with $N=4941$ nodes and $L=6594$ edges; a node indicates a generator or a transformator or a substation, and each edge represents a power supply line. 
We shall find it convenient to work here with the average degree $\langle k \rangle$ defined in Eq.\ (\ref{eq:kvsp}) (and we recall that the percolation phase transition of an ER random network would take place at $\langle k \rangle=1$).
The network, displayed in the top left panel of Fig.~\ref{fig:US_power_grid_network}, is made up of a single connected component and is characterized by an average degree $\langle k \rangle = 2.67$; its \emph{fill}, defined as the ratio between the number of edges $L$ and the maximum number of edges $N(N-1)/2$ in an undirected network without loops, takes the value $5.40\cdot10^{-4}$.

The US power grid network is deterministic, with the edges corresponding to an organized structure, constrained by infrastructural requirements. 
In order to enable a comparison with a random graph, we shall introduce randomness in the system, by sampling out a subset of $N'<N$ nodes to form an subgraph, in which the edges connecting these nodes are inherited from the original network.
The sub-networks corresponding to different numbers $N'$ of sub-sampled nodes are reported in Fig. \ref{fig:US_power_grid_network}.

\begin{figure}
\centering
\includegraphics[width=0.5\textwidth]{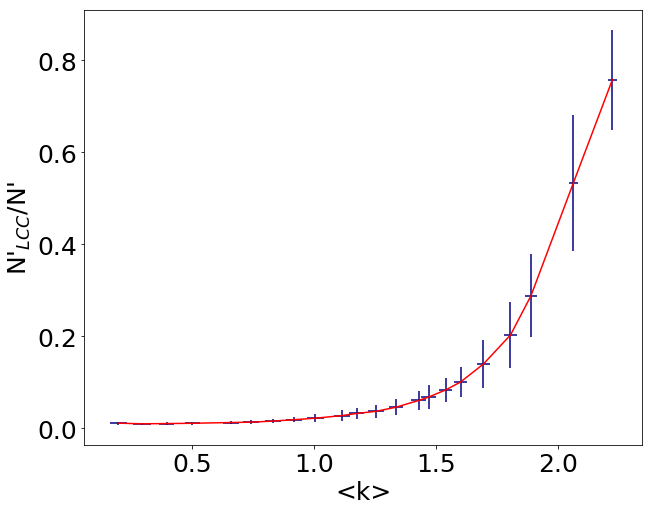}
\caption{Largest connected component $N'_{LCC}/N'$ vs average degree $\langle k \rangle$ in the sampled subnetworks. Each point corresponds to a fixed subsampling size $N'$: the coordinates represent the mean values obtained from $M=100$ realizations and the error bars coincide with the standard deviations of the respective distributions. The dependence is smooth and no signature of a phase transition is present.}
\label{fig:LCC_vs_kmean}
\end{figure}

\begin{figure}
\centering
\includegraphics[width=0.4\textwidth]{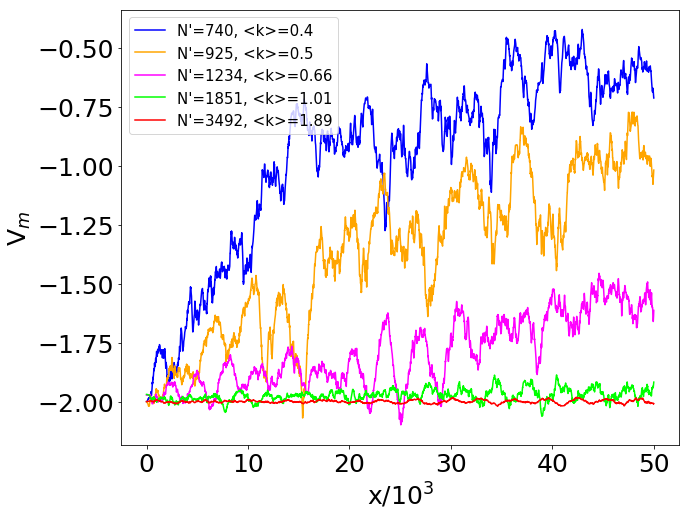}
\hspace{0.2cm}
\includegraphics[width=0.4\textwidth]{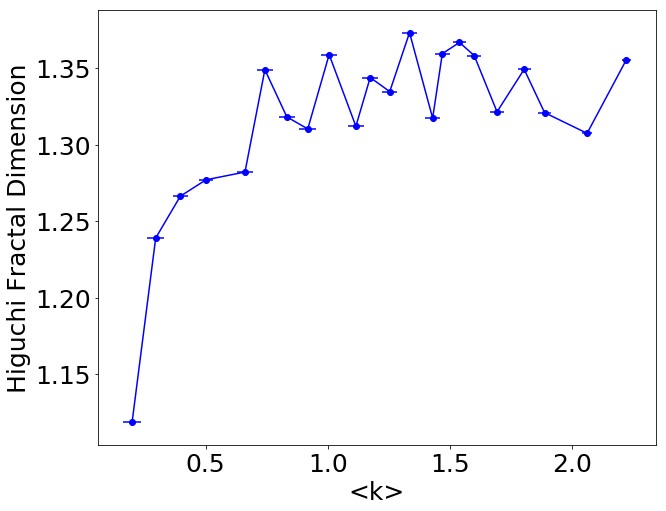}
\caption{Left. Pointwise median $V_m(x)$ of the reconstructed potentials of $M=100$ subsamplings of the US power grid with $N'$ nodes and edges inherited from the original network ($N=4941$). The size $N'$ of sampled subnetworks ranges from $N'=740$ (top) to $N'=3492$ (bottom).
Right. Higuchi fractal dimension of the median potentials on a set of $M=100$ subsamplings of the US power grid network, as a function of the average degree $\langle k \rangle$, for $370 \leq N' \leq 4113$. The horizontal coordinates indicate the mean of $\langle k \rangle$ on the ensemble of subsamplings, while the error bars denote the corresponding standard deviations.
}
\label{fig:V_median_undersampling}
\end{figure}

The properties of the sampled subnetworks will fluctuate according to the specific subset of $N'$ nodes. We performed a statistical analysis in the intermediate range $370\leq N' \leq 4113$, in which the variability of the sampled subnetwork is expected to be larger, as $0.20 \leq \langle k \rangle \leq 2.22$ would be close to criticality ($\langle k \rangle = 1$) if the network were random. 
At fixed $N'$, we generated $M=100$ subgraph realizations and computed, for each of them, the average degree $\langle k \rangle$, the size $N'_{LCC}$ of the largest connected component (LCC) and the reconstructed potential. 
Figure \ref{fig:LCC_vs_kmean} displays the dependence of  
the LCC relative size $N'_{LCC}/N'$ vs the average degree $\langle k \rangle$ of such distributions, with the error bars representing their standard deviations over $M=100$ subgraph realizations (at fixed $N'$). 
The dependence of the size of the LCC on $\langle k \rangle$ appears to be smooth and one observes no signature of a phase transition. We stress again that the graph is not random and a phase transition is not properly defined in this case.
Although this is not central to our analysis, we also observe that the number of subnetworks scales like $N \choose N'$, so that the 100 realizations are practically independent and very different from each other. 

We now test the performance of the reconstructed potential formalism as a tool of network analysis. 
In Fig.~\ref{fig:V_median_undersampling} (left) we show the profile of the pointwise median potential $V_m$ for some values of $N'$; the corresponding mean values of $\langle k \rangle$ are also reported to facilitate comparison with the analogous plots for the ER networks. Finally, we computed the HFD of the median potentials, represented in Fig.~\ref{fig:V_median_undersampling} (right) as a function of the average degree in the sampled subnetworks. 
Notice that it is impossible to investigate the behavior of the HFD for larger values of 
$\langle k \rangle$, as the US power grid  network, displayed in the top left panel of Fig.\ \ref{fig:US_power_grid_network}, has an average degree $\langle k \rangle\simeq 2.67$, that represents an upper bound for our analysis. 

The range of $\langle k \rangle$ values here analyzed roughly corresponds to probabilities $4 \times 10^{-4} \leq p \leq 4.4 \times 10^{-3}$ in Fig.\ \ref{fig:HFD_vs_p} (random ER networks). The HFD displays a clear increasing trend up to $\langle k \rangle\simeq 0.75$, and then reaches a plateau for larger values of $\langle k \rangle$, characterized by irregular (a few percent) fluctuations. 
Notice that, as emphasized before, the real US power grid network is characterized by an average degree $\langle k \rangle\simeq 2.67$, a value that cannot be reached in our analysis. Comparison with a \emph{bona fide} random ER network would lead us to expect that the plateau ends at 
$\langle k \rangle \gtrsim 3.5$ ($p \gtrsim 7 \times 10^{-3}$). On a wider scale the plateau would then appear as a (sharp) maximum, see Fig.\ \ref{fig:HFD_vs_p}. 

The behavior of the HFD detects therefore the occurrence of a significant amount of fractality in the ensemble of sampled subnetworks for the corresponding values of $N'$ and $\langle k \rangle$.
We can interpret this behavior as a remnant of the criticality of the corresponding ER graph. 
Observe that such conclusion does not emerge from the quantity displayed in Fig.~\ref{fig:LCC_vs_kmean}. The fractality of the potential, detected by the HFD, appears therefore to be a better indicator of a structure that is reminiscent of the ER phase transition.
It is remarkable that we are able to reach this conclusion although the sampled subnetworks originate from a given, real-world deterministic graph and cannot, strictly speaking, be considered ``random", as those of an ER graph.

A few additional comments are in order. We evaluated the HFD for $2\leq k \leq 800$, $k$ being the scale index introduced in Eq.\ (\ref{eq:lengthHFD}). This choise was motivated by consistency with the analysis of Sec.\ \ref{sec:higfract}, pertaining to an ER random network. 
An investigation with $2\leq k \leq 35$ would yield the same qualitative features, with a lower (about 30\%) fractality index, but a very similar plateau, starting again at $\langle k \rangle\simeq 0.75$.

\section{Conclusions and outlook}
\label{conclusions}
We have proposed a novel, quantum-inspired approach to investigate complex networks: by using the mathematical framework provided by dressing transformations, we have developed a technique to uniquely associate a Schr\"odinger-like potential to the graph spectrum of a given network. We have verified that such potential probes and detects nontrivial phenomena in complex network dynamics, such as the phase transition in the Erd\"{o}s-R\'{e}nyi model, related to the emergence of a giant component. In particular, such transition can be investigated by monitoring the length scale and the Higuchi Fractal Dimension of the median potential computed from several network realizations having the same size and connection probability. 

We have applied this technique to the study of a real-world network, showing that the fractality of the median potential displays a behavior that is reminescent of criticality, although no \emph{bona fide} phase transition can be properly defined in such a case. We also observed that standard techniques (such as the scaling feaures of the LCC) are unable to detect any signature or remnant of criticality. 
Future work will extend this analysis to more complex models of artificial networks, which include other formation mechanisms, such as link rewiring and preferential attachment. 

The reconstructed potential provides a snapshot representation of the structure of a network, yielding information on its connectivity and on the number of disconnected components. This aspect is particularly interesting in the perspective of examining real-world networks, since the reconstructed potential could be used to test their robustness, and even diagnose possible weaknesses. For this reason, we plan to explore further the characterization of real-world networks through reconstructed potentials, trying in particular to understand whether the typical patterns found in this analysis are an intrinsic feature of the specified domains. Moreover, we will investigate the possibility to improve our framework by combining the reconstructed potentials with other novel approaches to complex networks, based on entropy \cite{de2016spectral, monaco2018, monaco2019} and machine learning \cite{muscoloni2017}.

\section*{Supplementary Information}

\subsection*{Dressing transformations to reconstruct potentials from spectra}

We rapidly sketch the method we adopted for reconstructing a potential in a 1D Schr\"odinger equation and the corresponding eigenfunctions from a set of given energy levels.
Let us consider a one-dimensional symmetric potential $V(x)$, defining the Hamiltonian $H=p^2+V$, with $p=-\ii \partial_x$ (and $\hbar^2/2m=1$). Suppose that $E$ is the ground state energy of $H$. Let us now consider an arbitrary $\bar{E}<E$: since $\bar{E}$ cannot be an eigenvalue, the equation
\begin{equation}\label{Fx}
(p^2+V(x)) F(x) = \bar{E} F(x)
\end{equation}
can be solved only by non-normalizable functions. In particular, the equation admits a symmetric solution ($F'(0)=0$) with no nodes, whose inverse logarithmic derivative $f(x)=-F'(x)/F(x)$ satisfies the nonlinear equation
\begin{equation}
f'(x) - f^2(x) + V(x) = \bar{E}, \quad \text{with } f(0)=0.
\end{equation}
It is now possible to associate to $V$ a new potential $\bar{V}$, defined by
\begin{equation}
\bar{V}(x) = f'(x) + f^2(x) + \bar{E},
\end{equation}
and check that the function
\begin{equation}
\Psi(x) = \Psi(0) \exp\left( \int_0^x f(y) \dd y \right)
\end{equation}
is an eigenfunction of $\bar{H}=p^2+\bar{V}$ with eigenvalue $\bar{E}$. Moreover, being the inverse of the solution $F(x)$ of Eq.\ \eqref{Fx}, it is symmetric and has no node, hence corresponding to the ground state of $\bar{H}$.

The above properties can be iteratively used to build a potential $V_1(x)$ characterized by the given set of discrete energy levels $\{ E_n \}_{1\leq n \leq \widetilde{N}}$, with $E_n<E_{n+1}$, where $E_1$ is the ground state energy. The procedure can start from a constant potential $V_{\widetilde{N}+1}>E_{\widetilde{N}}$, so that the largest energy $E_{\widetilde{N}}$ falls below the (continuous) spectrum of $p^2+V_{\widetilde{N}+1}$. At each step, one determines the solution $f_n$ of
\begin{equation}\label{fn}
\left\{ \begin{array}{l} \displaystyle f'_n(x) - f_n^2(x) + V_{n+1}(x) - E_n = 0 \\ \\ f_n (0)=0 \end{array} \right.
\end{equation}
and then updates the potential as
\begin{equation}\label{Vn}
V_n(x) = f_n'(x) + f_n^2(x) + E_n = V_{n+1}(x) + 2 f_n'(x).
\end{equation}
It is also possible to verify that 
\begin{equation}\label{psi0n}
\psi_1^{(n)}(x) = \psi_1^{(n)}(0) \exp\left( \int_0^x f_n(y) \dd y \right) \quad \Rightarrow \quad (p^2+V_n(x))\psi_1^{(n)}(x) = E_n \psi_1^{(n)}(x) .
\end{equation}
Since, by construction, $E_n$ is below the spectrum of $p^2+V_{n+1}$, the function $f_n$ that satisfies \eqref{fn} yields a normalizable $\psi_1^{(n)}$ with no nodes, which is thus the ground state of $p^2+V_n$. The iteration proceeds until reaching $V_1(x)$, for which the lowest energy level is $E_1$, with eigenfunction $\psi_1^{(1)}$.

We are now ready to show that all the $E_n$'s are also energy levels of $V_1$, and construct the corresponding eigenfunctions. The fundamental result is that, considering the relations \eqref{Vn} and the commutator $[p,g(x)]=-\ii g'(x)$, 
\begin{equation}\label{commut}
(p^2+V_n)(p-\ii f_n) = (p-\ii f_n)(p^2+V_{n+1}).
\end{equation}
Due this relation, it is possible to verify that the normalizable wavefunctions
\begin{equation}
\psi_n^{(1)}(x) = \left[\prod_{j=1}^{n-1} (p-\ii f_j(x))\right] \psi_1^{(n)}(x) \propto \left[ \prod_{j=1}^{n-1} (\partial_x + f_j(x)) \right] \exp\left( \int_0^x f_j(y) \dd y \right)
\end{equation}
with $n>1$ and $\psi_1^{(n)}$ defined as in \eqref{psi0n}, satisfy
\begin{equation}
(p^2 + V_1(x)) \psi_n^{(1)}(x) = E_n \psi_n^{(1)}(x).
\end{equation}

\subsection*{Additional properties of reconstructed potential}

\begin{figure}
\includegraphics[scale=0.33]{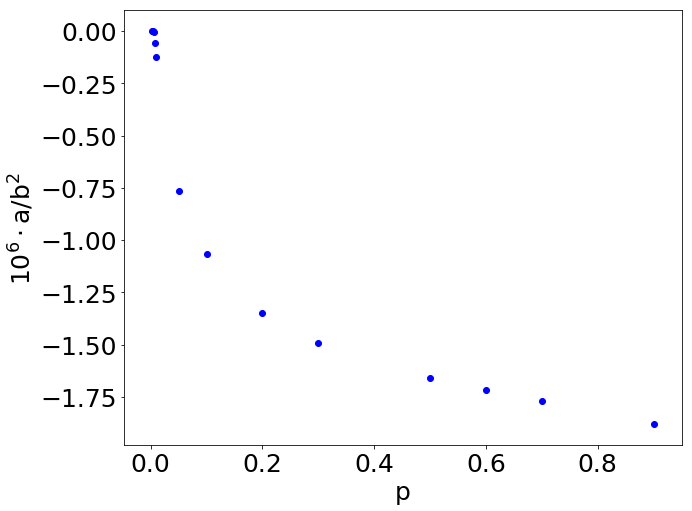}
\hspace{0.2cm}
\includegraphics[scale=0.33]{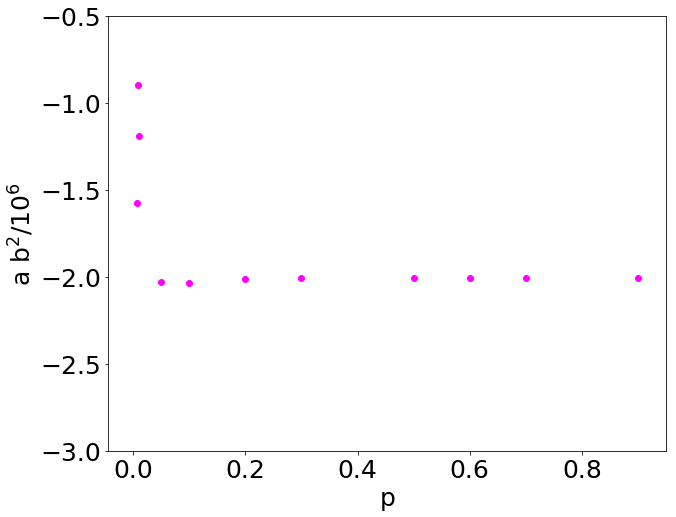}
\caption{Left: $a/b^{2}$ as a function of $p$. Right: $a b^{2}$, as a function of $p\geq 6\cdot 10^{-3}$.
}
\label{fig:ab-2}
\end{figure}

We add here a few comments on the structure of the Schrodinger potential associated with the Laplacian spectrum.
If we expand the potential 
\begin{equation}
\label{eq:Vfit}
    V_\textrm{fit}(x,p)= a(p)\,\mathrm{sech}^2\left(\frac{x}{b(p)} \right)-2-a(p)\,,
\end{equation}
for $x \ll b$, we obtain
\begin{equation}
\label{eq:Vab}
    V_\textrm{fit}(x,p)= -2 - \frac{a}{b^2} x^2 .
\end{equation}
The ration $a/b^2$ is therefore the concavity of the quadratic approximation (parabola) of the potential in the origin. 
Figure \ref{fig:ab-2} (left) displays the behavior of this quantity vs $p$. Notice that it tends to vanish for $p\to 0$, as expected from Eq.\ (9) of the main text if $2 \beta - \alpha> 0$. Our fit yields $2 \beta - \alpha \simeq 1$.

In Fig.\ \ref{fig:ab-2} (right) we plot the quantity $a b^2$ vs $p$. This quantity is proportional to $\int_0^\infty x V(x) dx$ (in general, $\int_0^\infty x^n V(x) dx \propto ab^{n+1}$).
Interestingly, it is almost constant for large values of $p$ (far from the phase trasition). For smaller $p$ it becomes very unstable (points 
$p < 6\cdot 10^{-3}$
not shown in the figure) and does not enable one to draw any solid conclusion.

Finally, we also observe that the functional form (\ref{eq:Vfit}) of $V$ could be valid also at $p<p_c$. However, the presence of oscillations and the roughness of the median potential make the fit of $a$ and $b$ very difficult $p<p_c$.

\subsection*{Numerical procedure}

In order to obtain $M$ realizations of the ER networks at critical connection probability for all the considered values of $N$, it was necessary to split the computation on different servers, in order to reduce the computational time. This task was made possible by implementing parallel computing on the infrastructures provided by the Bari ReCaS Datacenter. For example, $M=20000$ realizations of the ER networks with $N=500$ nodes were analyzed by employing $400$ different cores, each dedicated to the computation of reconstructed potentials related to $50$ networks. On each core, the task was accomplished in about $24$ hours: the same operations, run in series on a single device, would have required a $400$ times longer computational time. All the reconstructed potentials associated to the $20000$ random network realizations provided about $24$ Gigabyte (Gb) data, that were subsequently transferred to a single server, to compute the median potential and its fractal dimension, yielding the results shown in Fig.\ 11 of the main text. The value of $24$ Gb represents a limiting size for the amount of data that can be stored and processed simultaneously thru the 16 Gb RAM memory available on the single device. Therefore, it was not possible to extend the range of considered values of $M$ and $N$, and make solid statements on the fractal dimension in the limits $M,N \to \infty$. The data show nonetheless that the fractality of the median potential detects the percolation phase transition. 

\bibliography{biblio}

\section*{Acknowledgements}

Code development/testing and results were obtained on the IT resources hosted at ReCas data center. ReCaS is a project financed by the italian MIUR (PONa3$\_$00052, Avviso 254/Ric.). 
SP acknowledges support by MIUR via PRIN 2017 (Progetto di Ricerca di Interesse Nazionale), project QUSHIP (2017SRNBRK) and by INFN through the project ``QUANTUM".

\section*{Author contributions statement}

N.A. designed the study; N.A. and L.B. performed analyses and wrote the paper; S.P. supervised the development of the model and wrote the paper; R.B. supervised the research project and activity. All authors interpreted the results, revised the text and approved the final version of the paper.

\section*{Additional information}

\textbf{Competing interests}: The authors declare no competing interests.

\section*{Data availability}
The datasets generated and analysed during the current study are available from the corresponding author on reasonable request.

\end{document}